# Diffusion Waves in Sub-Quantum Thermodynamics:

# Resolution of Einstein's 'Particle-in-a-box' Objection

Gerhard Grössing,

*Austrian Institute for Nonlinear Studies,*

Akademiehof,

Friedrichstrasse 10, A-1010 Vienna, Austria

e-mail: ains@chello.at

**Abstract:** Einstein's objection against both the completeness claim of the orthodox version and the Bohmian interpretation of quantum theory, using the example of a "particle in a box", is reiterated and resolved. This is done by proving that the corresponding quantum mechanical states exactly match classical analogues. The latter are shown to result from the recently elaborated physics of diffusion waves.

# 1. Introduction: Einstein's objection

In 1953, Albert Einstein [1] summarized his arguments against the claim of the completeness of quantum theory, and also his criticism of Bohm's interpretation, by referring to the one-dimensional quantum mechanical problem of a particle of mass $m$ being trapped between two totally-reflecting walls of a box of length $L$. With quantum mechanics being a universal theory, Einstein argued, it should in principle also apply to macroscopic objects. Thus, the solution of the respective quantum mechanical problem should, at least approximately, approach the classical one when passing to the "macroscopic limit", like, e.g., when having a sphere of mass $m$ and diameter 1mm entrapped in a box of length 1m. As is well known, in the latter case there should be a classical to-and-fro uniform motion between the walls, which is something that not all quantum predictions do converge to.

More specifically, the quantum version of the problem is ideally represented by the location of two impenetrable walls at $x=0$ and $x=L$, with an external potential given by

$$V = \begin{cases} 0, & 0 < x < L \\ \infty, & x \leq 0, x \geq L \end{cases} \tag{1.1}$$

With the boundary conditions of $\psi = 0$ at $x = 0$ and $x = L$, one obtains the quantum mechanical wavefunctions of the normalized stationary state inside the box as

$$\psi_n(x,t) = \sqrt{\frac{2}{L}} \sin(k_n x) e^{-iE_n t/\hbar}, \tag{1.2}$$

where

$$k_n = \frac{n\pi}{L} \quad \text{and} \quad E_n = \hbar \omega_n = \frac{\hbar^2 k_n^2}{2m}. \tag{1.3}$$

With $p_n = \hbar k_n$, Fourier analysis of (1.2) provides a continuous distribution of momentum values $p$,

$$|\varphi_n(p)|^2 = \frac{\hbar}{8\pi^2 L}\left|\frac{-e^{-\frac{i}{\hbar}(p_n+p)L}}{p_n+p} - \frac{e^{\frac{i}{\hbar}(p_n-p)L}}{p_n-p} + \frac{2p_n}{p_n^2-p^2}\right|^2, \qquad (1.4)$$

which for high quantum numbers $n$ turns into the sum of two non-overlapping wavepackets peaked around the classical momenta $\pm p_n$, i.e.,

$$\lim_{n\to\infty}|\varphi_n(p)|^2 = \frac{1}{2}\left[\delta(p+p_n)+\delta(p-p_n)\right], \qquad (1.5)$$

just like for the classical case. Whereas Einstein considers this part of the theory as "completely satisfactory", he points to a severe problem when turning to possible position measurements. For, as can immediately be seen from (1.2), the particle position in the box is distributed as

$$|\psi_n(x,t)|^2 = \frac{2}{L}|\sin(k_n x)|^2, \quad 0 \leq x \leq L. \qquad (1.6)$$

Thus, due to the set of nodes with $|\psi_n(x,t)|^2 = 0$ along the length of the box, in certain points the particle can never be found. This is clearly incompatible with the classical to-and-fro motion mentioned above. But this means that the predictions of quantum theory do, at least in the case of high $n$ position measurements, not converge on the classical ones. This lead Einstein to conclude that quantum mechanics can at best be interpreted only as a statistical theory (i.e., via Born's rule), but is silent on the physics of individual particles. In other words, he considered quantum theory to be incomplete. Along with this objection, Einstein also expressed his criticism of Bohm's interpretation of quantum theory. For, the $x$-independence of the phase factor in (1.2) provides in the Bohmian picture that the velocity of the particle vanishes identically,

$$v = \frac{\nabla S}{m} = 0, \tag{1.7}$$

such that the particle's position becomes fixed at any nonnodal point along $L$. As a consequence, the kinetic energy is zero, and all the energy is contained in the "quantum potential" term $U$,

$$E_n = U := -\frac{\hbar}{2m}\frac{\nabla^2 |\psi_n|}{|\psi_n|} = \frac{p_n^2}{2m} = \frac{n^2\pi^2\hbar^2}{2mL^2}. \tag{1.8}$$

So, we have the contradiction that in the quantum case the particle is at rest, whereas classically it should have a momentum distribution as in (1.5) – a fact that Einstein criticized thus: "The disappearance of the velocity … contradicts the well-founded requirement that in the case of a macro-system the movement should approximately coincide with the one following from classical mechanics." [1]

Thus, one can pose the following question, which we call here *question A*:

*"How can the state of rest implied by the quantum theory of motion be compatible with the finite classical values obtained in a measurement of momentum?"* [2]

Peter Holland, in [2], gives a detailed account of the "particle in a box" problem, and also provides a consistent explanation within the causal (Bohmian) interpretation, which apparently answers Einstein's critique fully. He considers the causal version of said classical momentum measurement using the "time-of-flight" method: the confining walls of the box are suddenly removed, and the particle is detected at some point $x$ at a time $t \to \infty$, thus providing with the number $m\frac{x}{t}$ the value of the momentum immediately prior to the position measurement, with the position distribution echoing the momentum distribution (1.4). Then, two identical separating wavepackets can be seen to form, as the wavefunction at time $t$ after removal of the walls is

$$\psi_n(x,t) = \left(\frac{h^{1/2}}{2\pi}\right) \int_{-\infty}^{\infty} \varphi(\hbar k)\, e^{i\left(kx - \hbar k^2 t/2m\right)} dk, \quad \forall n. \tag{1.9}$$

Thus, starting from rest, the quantum potential energy $U$ is liberated with the spreading of the wave, and the particle gradually acquires kinetic energy, finally leading for $n \to \infty$ to the classical final momenta $\pm p_n$. Holland's resolution of the problem stresses the claim that the state (1.2) has no classical analogue, thus apparently providing a counter-example to what Einstein considered a "well-founded requirement", i.e., that classical mechanics should emerge in the "macroscopic limit" for *all* valid quantum mechanical states.

However, in the present paper an exact classical analogue to the state (1.2) will be derived, thus showing that Einstein's requirement can still be considered well-founded. The derivation will be based on the recently proposed foundation of quantum mechanics in nonequilibrium thermodynamics [3], which in essential ways goes beyond the Bohmian approach to a quantum theory of motion. This can immediately be made clear by "reverting" the above-quoted *question A*, which Holland had posed and answered within a Bohmian framework.

Thus, one can pose the following *question B*: *How can the classical momentum of the particle be compatible with the nodal structure of the distribution obtained in a quantum mechanical position measurement?* In other words, we consider the "reverse" experiment of suddenly capturing a previously free classical particle within the confining walls, and ask: *How can one understand that the kinetic energy converts completely into the energy of the quantum potential (which does provide said nodal structure)?* This question cannot be answered satisfactorily within a purely Bohmian approach, because one would thereby also have to answer the question of how the quantum mechanical wavefunction $\psi$, and with it, in consequence, the quantum potential $U$, appears, although we started with a purely classical particle. The reason why a purely Bohmian approach cannot answer *question B* satisfyingly, lies, of course, in the fact that in this approach, like in orthodox quantum theory, $\psi$ is not derived from an underlying theory, such that its appearance is not explained in any dynamical theory. As in reference [3], however, such a derivation was achieved, we can now turn to the respective dynamical theory to try to answer *question B*, i.e., to thermodynamics.

Therefore, in Section 2, a short recapitulation of reference [3] is given. Then, in Section 3, the starting point of our observations will be the simple fact that there is one thing that, along a single path, a free classical particle and a free quantum mechanical particle have in common: a vanishing quantum potential. As the latter in the thermodynamic framework does have a distinct classical meaning, one can easily show that its vanishing leads to a classical heat equation. Moreover, for a non-vanishing quantum potential, like in the "particle-in-a-box" problem, a dynamical equation is derived in Section 4, whose solution exactly matches the quantum state (1.2). The equation itself turns out as a type of pseudo-wave Helmholtz equation, exactly identical to the one describing classical diffusion-wave fields. In Section 5, then, the resolution of Einstein's objection is recapitulated with the aid of the previously introduced [3] vacuum fluctuation theorem. Finally, in Section 6, the results are summarized and some future perspectives are discussed.

## 2. Short resumée of the derivation of the Schrödinger equation from nonequilibrium thermodynamics

In reference [3], the fact that each particle of nature is attributed an energy $E = \hbar\omega$ is considered as central for the understanding of quantum theory. As oscillations, characterized by some typical angular frequency $\omega$, are today considered as properties of off-equilibrium steady-state systems, one can generally assume that "particles" are actually dissipative systems maintained in a nonequilibrium steady-state by a permanent throughput of energy, or heat flow, respectively. The latter must be expressed by some form of kinetic energy, which is not identical with the kinetic energy of the particle, but an additional (external) contribution to it. One can thus write down the energy of the total system, i.e., the particle as the "system of interest" and the heat flow as the particle's thermal context (*Assumption 1*):

$$E_{\text{tot}} = \hbar\omega + \frac{(\delta p)^2}{2m}, \tag{2.1}$$

where $\delta p$ is the additional, fluctuating momentum component of the particle of mass $m$.

Moreover, as has been suggested since the beginnings of quantum theory, the particle's environment may be considered such as to provide detection probability distributions which can be modelled by wave-like intensity distributions. Thus, the detection probability density $P(x,t)$ is considered to coincide with a classical wave's intensity $I(x,t) = R^2(x,t)$, with $R(x,t)$ being the wave's real-valued amplitude (*Assumption 2*):

$$P(\mathbf{x},t) = R^2(\mathbf{x},t), \text{ with normalization } \int P\, d^n x = 1. \tag{2.2}$$

In the present paper, there is no room to cover the fascinating results which throughout recent years have been accumulated in the field of nonequilibrium thermodynamics. (For an excellent review, see, e.g., ref. [4].) In ref. [3], we proposed to merge some results of said field with classical wave mechanics in such a way that the many microscopic degrees of freedom associated with the hypothesized subquantum medium can be recast into the more "macroscopic" properties that characterize the wave-like behaviour on the quantum level. Thus, the relevant description of the "system of interest" no longer needs the full phase space information of all "microscopic" entities, but needs only the "emergent" particle coordinates.

As we consider a particle as being surrounded by a "heat bath", i.e., a reservoir that is very large compared to the small dissipative system, one can safely assume that the momentum distribution in this region is given by the usual Maxwell-Boltzmann distribution – a wide-spread characteristic for similar systems, despite the fact that we deal with nonequilibrium thermodynamics. This corresponds to a "thermostatic" regulation of the reservoir's temperature, which is equivalent to the statement that the energy lost to the thermostat can be regarded as heat. Thus, without further delving into the matter more deeply here, one arrives via a corresponding *proposition of emergence* at the equilibrium-type probability (density) ratio (*Assumption 3*)

$$\frac{P(\mathbf{x},t)}{P(\mathbf{x},0)} = e^{-\frac{\Delta Q}{kT}}, \tag{2.3}$$

with $k$ being Boltzmann's constant, $T$ the reservoir temperature, and $\Delta Q$ the heat that is exchanged between the particle and its environment.

Equations (2.1), (2.2), and (2.3) are the only assumptions necessary to derive the Schrödinger equation from (modern) classical mechanics. In fact, we need only two additional well-known observations in order to achieve that goal. The first is given by Boltzmann's formula for the slow transformation of a periodic motion (with period $\tau = 2\pi/\omega$) upon application of a heat transfer $\Delta Q$. With the action function $S = \int (E_{kin} - V) dt$, the relation between heat and action is given as

$$\Delta Q = 2\omega \delta S = 2\omega \left[ \delta S(t) - \delta S(0) \right]. \tag{2.4}$$

Finally, as the kinetic energy of the thermostat is given by $kT/2$ per degree of freedom, and as the average kinetic energy of the oscillator is equal to half of the total energy $E = \hbar\omega$, the requirement that the average kinetic energy of the thermostat equals the average kinetic energy of the oscillator reads, for each degree of freedom, as

$$\frac{kT}{2} = \frac{\hbar\omega}{2}. \tag{2.5}$$

Now, combining the Equs. (2.3), (2.4), and (2.5), one obtains

$$P(\mathbf{x},t) = P(\mathbf{x},0) e^{-\frac{2}{\hbar} \left[ \delta S(\mathbf{x},t) - \delta S(\mathbf{x},0) \right]}, \tag{2.6}$$

from which follows the expression for the momentum fluctuation $\delta \mathbf{p}$ of (2.1) as

$$\delta \mathbf{p}(\mathbf{x},t) = \nabla \left( \delta S(\mathbf{x},t) \right) = -\frac{\hbar}{2} \frac{\nabla P(\mathbf{x},t)}{P(\mathbf{x},t)}. \tag{2.7}$$

This, then, provides the additional kinetic energy term as

$$\delta E_{kin} = \frac{1}{2m} \nabla(\delta S) \cdot \nabla(\delta S) = \frac{1}{2m} \left( \frac{\hbar}{2} \frac{\nabla P}{P} \right)^2. \tag{2.8}$$

Thus, writing down a classical action integral including this new term yields (with external potential $V$)

$$A = \int L\, d^n x dt = \int P(\mathbf{x},t)\left[\frac{\partial S}{\partial t} + \frac{1}{2m}\nabla S\cdot\nabla S + \frac{1}{2m}\left(\frac{\hbar}{2}\frac{\nabla P}{P}\right)^2 + V\right]d^n x dt. \qquad (2.9)$$

Introducing the "Madelung transformation"

$$\psi = Re^{\frac{i}{\hbar}S}, \qquad (2.10)$$

where $R = \sqrt{P}$ as in (2.2), one has, with bars denoting averages,

$$\overline{\left|\frac{\nabla\psi}{\psi}\right|^2} := \int d^n x dt \left|\frac{\nabla\psi}{\psi}\right|^2 = \overline{\left(\frac{1}{2}\frac{\nabla P}{P}\right)^2} + \overline{\left(\frac{\nabla S}{\hbar}\right)^2}, \qquad (2.11)$$

and one can rewrite (2.9) as

$$A = \int L dt = \int d^n x dt \left[|\psi|^2\left(\frac{\partial S}{\partial t} + V\right) + \frac{\hbar^2}{2m}|\nabla\psi|^2\right]. \qquad (2.12)$$

Thus, with the identity $|\psi|^2 \frac{\partial S}{\partial t} = -\frac{i\hbar}{2}\left(\psi^*\dot\psi - \dot\psi^*\psi\right)$, one obtains the familiar Lagrange density

$$L = -\frac{i\hbar}{2}\left(\psi^*\dot\psi - \dot\psi^*\psi\right) + \frac{\hbar^2}{2m}\nabla\psi\cdot\nabla\psi^* + V\psi^*\psi, \qquad (2.13)$$

from which by the usual procedures one arrives at the Schrödinger equation

$$i\hbar\frac{\partial\psi}{\partial t} = \left(-\frac{\hbar^2}{2m}\nabla^2 + V\right)\psi. \qquad (2.14)$$

Note also that from (2.9) one obtains upon variation in $P$ the modified Hamilton-Jacobi equation familiar from the de Broglie-Bohm interpretation, i.e.,

$$\frac{\partial S}{\partial t} + \frac{(\nabla S)^2}{2m} + V + U = 0, \qquad (2.15)$$

where $U$ is known as the "quantum potential"

$$U = \frac{\hbar^2}{4m}\left[\frac{1}{2}\left(\frac{\nabla P}{P}\right)^2 - \frac{\nabla^2 P}{P}\right] = -\frac{\hbar^2}{2m}\frac{\nabla^2 R}{R}. \tag{2.16}$$

Moreover, with the definitions

$$\mathbf{u} := \frac{\delta \mathbf{p}}{m} = -\frac{\hbar}{2m}\frac{\nabla P}{P} \text{ and } \mathbf{k_u} = -\frac{1}{2}\frac{\nabla P}{P} = -\frac{\nabla R}{R}, \tag{2.17}$$

one can rewrite $U$ as

$$U = \frac{m\mathbf{u}\cdot\mathbf{u}}{2} - \frac{\hbar}{2}(\nabla\cdot\mathbf{u}) = \frac{\hbar^2}{2m}(\mathbf{k_u}\cdot\mathbf{k_u} - \nabla\cdot\mathbf{k_u}). \tag{2.18}$$

(Note that there is a sign error in Equ. (3.2.29) of ref. [3].) However, as was already pointed out in ref. [3], with the aid of (2.4) and (2.6), $\mathbf{u}$ can also be written as

$$\mathbf{u} = \frac{1}{2\omega m}\nabla Q, \tag{2.19}$$

which thus explicitly shows its dependence on the spatial behavior of the heat flow $\delta Q$.

3. **The case of a vanishing quantum potential: Equivalence with the classical heat equation**

The energetic scenario of a steady-state oscillator in nonequilibrium thermodynamics is given by a throughput of heat, i.e., a kinetic energy at the subquantum level providing a) the necessary energy to maintain a constant oscillation frequency $\omega$, and b) some excess kinetic energy resulting in a fluctuating momentum contribution $\delta\mathbf{p}$ to the momentum $\mathbf{p}$ of the particle. From a perspective out of everyday life, one can compare this to the situation of some small convex half-sphere, say, lying on a flat vibrating membrane. Due to resonance, the half-sphere will oscillate with the same frequency as the membrane, but if the energy of the membrane's vibration is higher than that required for the half-sphere to co-oscillate, the latter will start to perform an irregular motion, thus reflecting minute irregularities in the membrane (or the half-sphere itself) such as to amplify them in a momentum fluctuation. However, there is one more element in the energy scenario that is important. In our everyday life example, it is the friction between the half-sphere and the membrane, which causes the half-sphere to dissipate heat energy into its environment.

Very similarly, the steady-state resonator representing a "particle" in a thermodynamic environment will not only receive kinetic energy from it, but, in order to balance the stochastic influence of the buffeting momentum fluctuations, it will also dissipate heat into the environment. In fact, the "vacuum fluctuation theorem" (VFT) introduced in ref. [3] proposes, as all fluctuation theorems, that the larger the energy fluctuation of the oscillating "system of interest" is, the higher is the probability that heat will be dissipated into the environment rather than be absorbed. Also, Bohm and Hiley [5] demand in their review of stochastic hidden variable models that, generally, to maintain an equilibrium density distribution like the one given by $P(x,t)$ under random processes, the latter *must* be complemented by a balancing movement. The corresponding balancing velocity is called, referring to the same expression in Einstein's work on Brownian motion, the "osmotic velocity". If we remind ourselves of the stochastic "forward" movement in our model, i.e., $\delta\mathbf{p}/m = \mathbf{u}$, or the current $\mathbf{J} = P\mathbf{u}$, respectively, this will have to be balanced by the osmotic velocity $-\mathbf{u}$, or $\mathbf{J} = -P\mathbf{u}$, respectively.

Inserting (2.17) into the definition of the "forward" diffusive current $\mathbf{J}$, and recalling the diffusivity $D = \hbar/2m$, one has

$$\mathbf{J} = P\mathbf{u} = -D\nabla P, \qquad (3.1)$$

which, when combined with the continuity equation $\dot{P} = -\nabla \cdot \mathbf{J}$, becomes

$$\frac{\partial P}{\partial t} = D\nabla^2 P. \qquad (3.2)$$

Equs. (3.1) and (3.2) are the first and second of Fick's laws of diffusion, respectively, and $\mathbf{J}$ is called the diffusion current.

So, whereas in ref. [3] we concentrated on that part of the energy throughput maintaining the particle's frequency $\omega$ that led to an additional momentum contribution $\delta p$ from the environment to be absorbed by the particle, we now are going to focus on the "other half" of the process, i.e., on the "osmotic" type of dissipation of energy from the particle to its environment. (The VFT, then, gives relative probabilities for the respective cases, to which we shall return below.)

Returning now to Equ. (2.19), and remembering the strict directionality of any heat flow, we can redefine this equation for the case of heat dissipation where $\Delta Q = Q(t) - Q(0) < 0$. Maintaining the heat flow as a positive quantity, i.e., in the sense of measuring the positive amount of heat dissipated into the environment, one therefore chooses the negative of the above expression, $-\Delta Q$, and inserts this into (2.19), to provide the osmotic velocity

$$\overline{\mathbf{u}} = -\mathbf{u} = D\frac{\nabla P}{P} = -\frac{1}{2\omega m}\nabla Q, \qquad (3.3)$$

and the osmotic current is correspondingly given by

$$\overline{\mathbf{J}} = P\overline{\mathbf{u}} = D\nabla P = -\frac{P}{2\omega m}\nabla Q. \qquad (3.4)$$

Then the corollary to Fick's second law becomes

$$\frac{\partial P}{\partial t} = -\nabla \cdot \overline{\mathbf{J}} = -D\nabla^2 P = \frac{1}{2\omega m}\left[\nabla P \cdot \nabla Q + P\nabla^2 Q\right]. \tag{3.5}$$

With these ingredients at hand, let us now return to the expression (2.18) for the quantum potential $U$, and let us see how we can understand its thermodynamic meaning. To start, we study the simplest case, which is nevertheless very interesting, i.e., $U = 0$. For, let us remind ourselves that we are interested in the similarities and differences between descriptions in the quantum and classical frameworks, respectively, and whether or not the descriptions within these two frameworks can be brought into full agreement. Certainly, one situation that is comparable for both the quantum and the classical descriptions is when we have to do with a free particle along a single path. Then, of course, both in the quantum and in the classical case, the quantum potential will vanish identically. (Remember at this point that in our thermodynamic ansatz, the quantum potential does have a "classical" meaning, too, as it actually was derived from purely classical physics.)

So, let us now consider $U = 0$, and then focus on the dynamics as we follow the behaviour of the osmotic velocity (3.3) that represents the heat dissipation from the particle into its environment. Firstly, we have from (2.18) that

$$\frac{\hbar}{2}(\nabla \cdot \mathbf{u}) = \frac{m\mathbf{u} \cdot \mathbf{u}}{2}. \tag{3.6}$$

Insertion of (2.19) yields the general thermodynamic corollary of a vanishing quantum potential as

$$\nabla^2 Q = \frac{1}{2\hbar\omega}(\nabla Q)^2. \tag{3.7}$$

However, turning now to the osmotic current and the flux behavior (3.5), we firstly insert (2.5) into (2.3) to give

$$P = P_0 e^{-\frac{\Delta Q}{\hbar \omega}}, \qquad (3.8)$$

and then obtain from (3.5) that

$$\frac{\partial P}{\partial t} = \frac{P}{2\omega m}\left[\nabla^2 Q - \frac{(\nabla Q)^2}{\hbar \omega}\right]. \qquad (3.9)$$

The last term on the r.h.s. can be rewritten with (3.7) to provide

$$\frac{\partial P}{\partial t} = -\frac{P}{2\omega m}\nabla^2 Q. \qquad (3.10)$$

Now, from (3.8) we also have

$$\frac{\partial P}{\partial t} = -\frac{P}{\hbar \omega}\frac{\partial Q}{\partial t}, \qquad (3.11)$$

so that comparison of (3.10) and (3.11) finally provides, with $\hbar\omega = \text{constant}$ and $\widetilde{Q} := Q/\hbar\omega$,

$$\nabla^2 \widetilde{Q} - \frac{1}{D}\frac{\partial \widetilde{Q}}{\partial t} = 0, \qquad (3.12)$$

or, generally,

$$\nabla^2 Q - \frac{1}{D}\frac{\partial Q}{\partial t} = 0. \qquad (3.13)$$

Equ. (3.13) is nothing but the *classical heat equation*, obtained here by the requirement that $U = 0$. In other words, *even for free particles*, both in the classical and in the quantum case, one can identify a *heat dissipation process emanating from the particle. A non-vanishing "quantum potential"*, then, is a means to describe the *spatial and temporal dependencies of the corresponding thermal flow in the case that the particle is not free.*

As to solutions to (3.12), let us consider some approximations at first. If, for example, we chose our temporal resolution such that, on average, the heat flow was constant, we would obtain from (3.12) a simple Laplace equation,

$$\nabla^2 \widetilde{Q} = 0, \tag{3.14}$$

whose solutions are harmonic functions, as, in three dimensions, and with $r = \sqrt{x^2 + y^2 + z^2}$, given by $\widetilde{Q} \propto \frac{1}{r}$, i.e., a singularity like a unit point charge at the origin. Further, if the temporal behaviour of $Q$ were not periodic (as we shall discuss lateron), but represented a heat generated only once and then dissipated, i.e., $Q \propto e^{-\omega t}$, then $\dot{Q} = -\omega Q = -Dk^2 Q$. Thus, with (3.12) one would then obtain the spatial Helmholtz equation

$$\left(\nabla^2 + k^2\right)\widetilde{Q} = 0 \tag{3.15}$$

with periodic solutions $\widetilde{Q} \propto e^{-i\mathbf{k}\cdot\mathbf{r}}$. Extending to the case of the inhomogeneous Helmholtz equation, consider the source as given by a delta function, i.e.,

$$\left(\nabla^2 + k^2\right)\widetilde{Q}(\mathbf{x}) = \delta(\mathbf{x}). \tag{3.16}$$

For a unique solution one introduces the Sommerfeld radiation condition, for example, as a specification of the boundary conditions at infinity. Then $\widetilde{Q}(\mathbf{x})$ is well known to be identical to a Green's function $G(\mathbf{x})$, which is given in three dimensions as

$$\widetilde{Q}(\mathbf{x}) = G(\mathbf{x}) = \frac{e^{ik|\mathbf{x}|}}{4\pi|\mathbf{x}|}. \tag{3.17}$$

So, we see that the heat equation (3.13) provides, even for the case that $U=0$, wave-like solutions $Q$, with the origin, or the particle's position, as source. In other words, one can identify solutions of the heat equation with Huygens-type waves.

Returning now to the problem of the particle in the box, we first observe the following. In order to accommodate time-periodic solutions of $Q$ (which are constitutive for our dissipative model, where the particle's frequency $\omega$ comes along with a periodic heat dissipation), one has to introduce a source term proportional to $e^{i\omega t}$ on the r.h.s. of (3.13), which, eventually, may turn out to vanish identically for $U=0$. The source term, of course, is chosen such that $\dot{Q}=i\omega Q$ does not fade out with time, as in the homogeneous Helmholtz case.

Now, for the particle-in-the-box problem, we begin with the "no box" situation, according to our attempt to answer *question B* from Section 1. The reverse scenario to not knowing in which direction the particle will leave the zone previously occupied by the "box" (i.e., to the left or to the right), is not knowing from which side it will enter the zone which is later to become that of the "box". This means that in an idealized experiment one can take a single-particle source, split up the path with a beam splitter, and later reflect the two separated paths in such a way as to confront them "head on" along a single "empty" line (i.e., one which will later be "filled" by the sudden insertion of two walls, thus confining the particle in a "box" then).

Looking at the "empty" line along which a particle either goes from left to right, or from right to left, we have two plane waves crossing each other. Whether we still deal with the case that $U = 0$, we shall have to find out. Thus, we write down the general ansatz

$$\nabla^2 \widetilde{Q} - \frac{1}{D}\frac{\partial \widetilde{Q}}{\partial t} = q(x)e^{i\omega t}. \tag{3.18}$$

Here, $q(x)$ is, for dimensionality requirements, proportional to $k^2$:

$$q(x) = q_1(x) + q_2(x) := -k^2 \widetilde{Q} e^{-i\omega t} + \overline{k}^2 \widetilde{Q} e^{-i\omega t}. \tag{3.19}$$

A first part, $q_1(x)$, is given in such a way that the generally time-independent part of (3.18) just covers the ordinary Helmholtz equation. The second part, however, is essentially given by the two classical plane waves representing (in an ensemble of many identical runs of the experiment) the two possible paths along which the particle may come with momentum $\pm \hbar k_0$, i.e.,

$$q_2(x) = \overline{k}^2 \widetilde{Q} e^{-i\omega t} := k_0^2 \cdot \frac{1}{2}\left[ e^{ik_0 x} + e^{-ik_0 x} \right]. \tag{3.20}$$

Note that if one performs a Fourier transformation of (3.20), one obtains that

$$q_2(k) = k_0^2 \cdot \frac{1}{2}\left[ \delta(k+k_0) + \delta(k-k_0) \right], \tag{3.21}$$

which is just the expression expected for the classical momentum distribution (1.5) and the free particle case. In the next Section, we shall "put this system into a box", and see how a classical theory can provide a continuous description of the transition from the situation of a free particle to that of a particle in a box.

Here, we just observe that with (3.20)

$$q_2(x) = k_0^2 \cos(k_0 x). \tag{3.22}$$

Insertion into (3.19) then provides, as

$$\widetilde{Q} = \frac{1}{k_0^2} q_2(x) e^{i\omega t} = \cos(k_0 x) e^{i\omega t}, \tag{3.23}$$

that

$$\nabla^2 \widetilde{Q} - \frac{1}{D}\frac{\partial \widetilde{Q}}{\partial t} = -k_0^2 \widetilde{Q} - \frac{i\omega}{D}\widetilde{Q} =: -\left(k_0^2 + \kappa^2\right)\widetilde{Q}. \qquad (3.24)$$

So, we find that we have to do with a non-vanishing quantum potential. In the next Section, the expression $\kappa^2 = i\omega/D$ will appear again, and we postpone its discussion until then.

In any case, we have seen that even in the situation of free particles there exists, at any time, *a non-vanishing undulatory heat flow* (3.23) *emanating from the particle*. This means that even a free particle is a permanent source of Huygens-type thermal waves, such that the superposition of two possible paths may already lead to a non-vanishing quantum potential. As a next example, we shall now look into the situation where such a free particle is, together with its thermal surroundings, entrapped between the walls of a confining box.

### 4. The case of a non-vanishing quantum potential: Equivalence with the classical thermal-wave-field equation

Considering the case $U \neq 0$, we now introduce an explicitly nonvanishing source term on the r.h.s. of (3.12), i.e.,

$$\nabla^2 \widetilde{Q} - \frac{1}{D}\frac{\partial \widetilde{Q}}{\partial t} = q(x)e^{i\omega t} \neq 0. \qquad (4.1)$$

Formally, one can solve this equation via separation of variables. Thus, with the ansatz

$$\widetilde{Q} = X(x)T(t), \quad \text{with} \quad T = e^{i\omega t}, \qquad (4.2)$$

one has $\nabla^2(XT) = T\nabla^2 X = \frac{1}{D} X \frac{\partial}{\partial t} T + q(x) T$. With

$$q(x) := \alpha(x) X, \tag{4.3}$$

this becomes $T\nabla^2 X = \frac{1}{D} X \frac{\partial}{\partial t} T + \alpha XT$. Division by $(XT)$ then provides the constant

$$\frac{\nabla^2 X}{X} = \frac{\frac{\partial}{\partial t} T}{DT} + \alpha = -\lambda. \tag{4.4}$$

Remembering that according to the construction the walls of our box are infinitely high, we can introduce the Dirichlet boundary conditions, i.e.,

$$\widetilde{Q}(0,t) = \widetilde{Q}(L,t) = 0, \tag{4.5}$$

which provides the constant $\lambda$ as

$$\lambda = \frac{n^2 \pi^2}{L^2} =: k_n^2. \tag{4.6}$$

With (4.4) one obtains, with normalization $\mathcal{N}$,

$$X = \mathcal{N} \sin\left(\frac{n\pi}{L} x\right), \tag{4.7}$$

and, furthermore, with $T = e^{i\omega_n t}$, $\frac{i\omega_n}{D} + \alpha = -k_n^2$, and therefore

$$\alpha = -k_n^2 (1+i). \tag{4.8}$$

With (4.3) one thus obtains

$$\mathcal{N} q(x) = -k_n^2 (1+i) \mathcal{N} \sin\left(\frac{n\pi}{L} x\right), \tag{4.9}$$

and, with (4.2) and (4.6), with $\widetilde{Q}_\mathcal{N} := \mathcal{N}\widetilde{Q}$,

$$\widetilde{Q}_\mathcal{N}(x,t) = \mathcal{N} \sin(k_n x) e^{i\omega_n t}. \tag{4.10}$$

Note that, due to the Dirichlet boundary conditions, $e_n := \mathcal{N} \sin(k_n x)$ are eigenvectors of the Laplacian,

$$\nabla^2 e_n = -k_n^2 e_n, \tag{4.11}$$

and

$$\langle e_n, e_m \rangle = \int e_n(x)\, e_m(x)\, dx = \begin{cases} 0 & m \neq n \\ 1 & m = n \end{cases}. \tag{4.12}$$

This means that, for $m = n$, (4.12) can be interpreted as a probability density, with

$$\int_0^L P\, dx = \mathcal{N}^2 \int_0^L \sin^2(k_n x)\, dx = 1. \tag{4.13}$$

The normalization thus derives from (4.13) as

$$1 = \mathcal{N}^2 \int_0^L \frac{1}{2}[1 - \cos(2k_n x)]\, dx = \mathcal{N}^2 \frac{L}{2},$$

and therefore

$$\mathcal{N} = \sqrt{\frac{2}{L}}. \tag{4.14}$$

So, we obtain the result that the heat distribution in the box is given by

$$\tilde{Q}_\mathcal{N}(x,t) = \sqrt{\frac{2}{L}} \sin(k_n x) e^{i\omega_n t}, \tag{4.15}$$

with the probability density

$$P = |\tilde{Q}_\mathcal{N}(x,t)|^2 = \left(\frac{L}{2}\right)^{-1} |\tilde{Q}(x,t)|^2. \tag{4.16}$$

Thus, the classical state (4.15) is shown to be identical with the quantum mechanical one, (1.2). Then, one obtains with (4.16) and (2.16) the quantum potential as

$$U = \frac{\hbar^2 k_n^2}{2m}. \tag{4.17}$$

This means also that the quantum potential now constitutes the total energy, and the kinetic energy of the particle, according to (4.15) and (1.7), vanishes identically.

Moreover, after obtaining the formal solution of (4.1), we can now also provide a more physical explanation. Similar to (3.19), we can again propose that

$$q(x) = q_1(x) + q_2(x) = -k^2 \tilde{Q} e^{-i\omega t} + \bar{k}^2 \tilde{Q} e^{-i\omega t} \neq 0. \tag{4.18}$$

As the particle has been trapped between the confining walls of the box, one has to change the expression on the r.h.s. of (3.20) to account for the fact that, relative to one particle path, the opposing one is a reflected one (i.e., off a wall), thus changing the phase by $e^{i\pi} = -1$. Therefore, our ansatz now becomes

$$q_2(x) = \overline{k}^2 \widetilde{Q} e^{-i\omega t} = k_n^2 \cdot \frac{1}{2}\left[e^{-ik_n x} - e^{ik_n x}\right]. \tag{4.19}$$

Incidentally, note that the r.h.s. of (4.19) corresponds to the Fourier-transformed expression

$$q_2(k) = k_n^2 \cdot \frac{1}{2}\left[\delta(k+k_n) - \delta(k-k_n)\right], \tag{4.20}$$

i.e., thereby also accounting for the reflection as opposed to the free case in (3.21).

As $\sin(k_n x) = \frac{1}{2i}\left(e^{ik_n x} - e^{-ik_n x}\right)$, we obtain from (4.19) with (4.18) that

$$\mathcal{N}q(x) = -(1+i)k_n^2 \mathcal{N} \sin(k_n x), \tag{4.21}$$

such that, just as in (3.24), one has an eigenvalue equation (dropping the index $\mathcal{N}$)

$$\nabla^2 \widetilde{Q} - \frac{1}{D}\frac{\partial \widetilde{Q}}{\partial t} = q(x)e^{i\omega t} = -(1+i)k_n^2 \widetilde{Q}, \tag{4.22}$$

with the unique solutions for $\widetilde{Q}$ now given by (4.15).

Finally, we can observe that by applying a temporal Fourier transformation on (4.22) and introducing the complex diffusion wave number $\kappa(x,\omega) := \sqrt{\frac{i\omega}{D}}$, one obtains a Helmholtz-type pseudo-wave equation:

$$\nabla^2 \widetilde{Q}(x,\omega) - \kappa^2 \widetilde{Q}(x,\omega) = Q(x,\omega). \tag{4.23}$$

Equ. (4.23), however, along with the identical definition of $\kappa$, is the exact *defining equation for a thermal-wave-field* and thus describes the spatio-temporal behaviour of *diffusion waves*. [6]

In his extensive survey of diffusion-wave fields [6], Andreas Mandelis stresses that equations of the type (4.23), or the heat equation, respectively, are peculiar in the sense that they are parabolic partial differential equations with no second-order time derivatives. Whereas for wavelike hyperbolic equations including the latter, families of solutions exist in terms of forward and backward waves propagating in space, no such solutions generally exist for parabolic equations, which are (mostly) deprived of the possibility of reflections at interfaces and of the existence of wave fronts, respectively. In contrast, they are characterized by an infinite speed of propagation of thermal disturbances along their entire domains. (Naturally, this feature makes them particularly amenable for modelling quantum mechanical nonlocality.)

Why, then, we must ask, is the probability density in our example of a particle entrapped between two walls given by (4.16), i.e., by an expression for standing waves constructed from the thermal waves (4.15) ? The answer is given by the specific physics of the initial conditions of our problem, i.e., by the two classical plane waves entering the area of the box, their relative phase difference due to reflection, and their superposition into standing waves. This leads to the heat equation (4.22) actually being an eigenvalue equation, with the sinusoidal expression of (4.15) as source. In other words, the source of the thermal waves is *already distributed* according to the standing wave pattern, and the continuation of the heat dissipation (i.e., the driving force) corresponds to *maintaining of the latter*. As with the construction of the standing waves nodes are generated, the particle will eventually become trapped between two such nodes and continually give off its momentum to the surrounding heat bath. Then, in the long time limit, it will have lost all its momentum to the heat bath, and the distribution of the latter will completely constitute that of the total energy in the form of the quantum potential. In the next Section, we shall see that some of the qualitative statements just made can be exactly formalized with the aid of the vacuum fluctuation theorem introduced in [3].

## 5. Resolution of Einstein's Objection and the Vacuum Fluctuation Theorem

We have seen that the quantum mechanical state (1.2) does have an exact classical analogue, i.e., (4.15). This is therefore in agreement with Einstein's "well-founded requirement" that in the "macroscopic limit" the movement of a classical object should emerge from the corresponding quantum mechanical one, at least approximately. However, the exact matching of (1.2) and (4.15) does not confirm Einstein's assertion that the "disappearance of the velocity" of the particle in a box contradicted said well-founded requirement. In fact, we have clearly seen why this is so: Even a free particle, be it quantum or classical, is a source of thermal waves. The latter are dissipated spherically-symmetrically from the particle position and have no net effect on particle motion as long as it is free. However, as soon as some kind of (classical) potential $V$ is operative, the configuration of the thermal waves in the new setting will change according to (4.23). Although the latter can be considered a classical equation, we have shown that it also corresponds exactly to a non-vanishing quantum potential. Therefore, even the "disappearance of the velocity" of a particle in a box can be described in classical terms, with an exact quantum mechanical analogue.

Moreover, one can now also make use of the "vacuum fluctuation theorem" (VFT) introduced in [3] to illustrate the problem of the particle in a box from a slightly different point of view. Consider the walls of our box initially being placed at the positions of plus and minus infinity, respectively, and let them approach each other to create a box of finite length $L$. Then we are dealing with a non-conservative system for which the average work $\overline{\Delta W}$ is given by [3]

$$\overline{\Delta W} = -\left\langle t \frac{\partial E}{\partial t} \right\rangle \frac{d\omega}{\omega}, \tag{5.1}$$

with brackets denoting time-averaging. As can easily be shown within classical thermodynamics, for the-particle-in-a-box problem this expression equals [7]

$$\overline{\Delta W} = -\mathbf{v} \cdot \mathbf{p} \frac{\delta L}{L} = -2 E_{\text{kin}} \frac{\delta L}{L}. \tag{5.2}$$

As can also be shown, this is identical to the quantum mechanical result: Writing the Schrödinger equation in terms of the Hamiltonian $H$, eigenfunctions $\psi_n(L)$, and eigenvalues $E_n(L)$,

$$\{H + V(L)\} \psi_n(L) = E_n(L) \psi_n(L), \tag{5.3}$$

we obtain after differentiation

$$\frac{\partial V}{\partial L} \psi_n + \{H + V\} \frac{\partial \psi_n}{\partial L} = \frac{\partial E_n}{\partial L} \psi_n + \left( E_n \frac{\partial \psi_n}{\partial L} \right). \tag{5.4}$$

As $\{H + V\}$ is hermitian, scalar multiplication with $\psi_m$ provides

$$\left( \frac{\partial V}{\partial L} \right)_{mn} = \frac{\partial E_n}{\partial L} \delta_{mn} + (E_n - E_m) \psi_m \frac{\partial \psi_n}{\partial L}. \tag{5.5}$$

Now, if the movement of the wall is so slow that the system's state is unchanged, i.e., that $m = n$, one obtains that although the energy $E_n \to E_n(L+\delta L)$, it is maintained in a quasi-stationary state. The change of the box length from $L$ to $L+\delta L$ is accompanied by a raising of the $n-$th energy level by the amount $\frac{\partial E_n}{\partial L}\delta L$, which is equal to the work applied to the system. [8] With $E_n$ given by $E_n = \frac{n^2\pi^2\hbar^2}{2mL^2}$, this amount of work equals

$$\Delta W = \frac{\partial E_n}{\partial L}\delta L = -\mathbf{v}\cdot\mathbf{p}\frac{\delta L}{L}, \quad (5.6)$$

which exactly matches (5.2).

Now let us formulate the VFT for our present purposes. Generally, it gives the probability ratio for heat dissipation versus heat absorption of a small object in a thermal bath with the aid of the dissipation function $\overline{\Omega_t}$. [4] In our case, it equals $\overline{\Omega_t}t = \overline{\Delta W}/kT$, with the identity of $kT = E_{\text{kin}}$ for the kinetic temperature of the heat reservoir [3]. The VFT then reads with (5.2) as

$$\frac{p(\overline{\Omega_t} = A)}{p(\overline{\Omega_t} = -A)} = e^{At} = e^{-2\frac{\delta L}{L}}, \quad (5.7)$$

where $p(A)$ is the probability for heat dissipation, and $p(-A)$ the probability for heat absorption. Now we can distinguish two cases, which have a direct correspondence to the situations discussed above with respect to *questions A and B*, respectively.

**a)** $L$ getting smaller; $\delta L < 0$: In the first case, we start with the two walls of the box far apart and moving towards each other such that the box length becomes continually smaller. Then, according to the VFT, the probability for heat absorption diminishes with increasing $|\delta L|$, i.e.,

$$p(-A) = e^{-2\left|\frac{\delta L}{L}\right|} p(A). \tag{5.8}$$

In other words, the particle subject to momentum fluctuations of the environment (due to the non-conservative system) will tend *not* to absorb an additional momentum fluctuation $\delta p$, but rather dissipate an amount of the kinetic energy in the form of heat. As this continues as long as the process goes on, one has in the long time limit that the particle's momentum $p$ tends towards zero. Conversely, we now consider case **b)** $L$ getting larger; $\delta L > 0$: Then the VFT provides

$$p(-A) = e^{2\frac{\delta L}{L}} p(A) = \left(1 + \frac{2\delta L}{L}\right) p(A). \tag{5.9}$$

In other words, the larger $\delta L$, the higher the probability becomes that the particle will absorb heat rather than dissipate it. Thus, as long as this process continues, the particle will acquire additional momentum, $\delta p$, due to the widening of the length $L$ of the box, or a moving apart of the walls, respectively. One could also say that the walls, in moving apart, tend to "drag" the particle with them, which is a valid illustration considering that the sinusoidal distribution of the probability density is maintained as long as the walls are separated by a finite distance. Then, of course, the nodes for each energy level $n$ must move accordingly, and therefore also any particle situated anywhere between two such nodes. Finally, with the walls infinitely far apart, the node structure will be gone and the particle will have absorbed all the heat energy (which is initially, as we have seen, identical with the energy due to the quantum potential) and thus move with the full kinetic energy $\pm p_n^2/2m$.

## 6. Conclusions and future perspectives

In this paper, we have continued to model quantum mechanics as emerging from a sub-quantum, and generally non-equilibrium, thermodynamics. Whereas in the previous paper [3] the exact Schrödinger equation was derived from a corresponding classical ansatz, we have here discussed some basic consequences. As in our approach the "particle's" frequency $\omega$ is conceived as the result of continuous energy throughput in an off-equilibrium steady state, we have concentrated in [3] on momentum fluctuations $\delta p$ *from the environment to the particle,* thereby obtaining deviations from the classical motion to provide exactly the quantum motion. In the present paper, on the other hand, we have focused on the "other half" of this process, i.e., on excess energy dissipated as heat *from the particle into its environment*. We found that this process is described by a form of the quantum potential $U$, even in the case of the free particle where $U=0$. In this latter case, the particle just emits spherically symmetrical thermal waves centered around its location, which have no net effect on the motion of the particle. However, as soon as this symmetry is broken, e.g., by the presence of some external potential $V$, the configuration of the thermal waves becomes distorted, with a direct effect on particle motion that is exactly equivalent to the effect of the quantum potential. However, with our new approach we were able to show that the "quantum potential" is essentially a way to describe said thermal configurations, and the latter are more aptly described by the physics of diffusion-wave fields, which have recently gained much attention in the literature [6]. Actually, taking into account the correct dimensionalities, one can now rewrite the usual modified (i.e., by the "quantum potential") Hamilton-Jacobi equation (2.15) as

$$\frac{\partial S}{\partial t}+\frac{(\nabla S)^2}{2m}+V-\frac{\hbar^2}{4m}\tilde{q}(x,k_u)e^{i\omega t}=0 \qquad (6.1)$$

or, equivalently, with $\tilde{Q}:=Q/\hbar\omega$, as

$$\frac{\partial S}{\partial t}+\frac{(\nabla S)^2}{2m}+V-\frac{\hbar^2}{4m}\left\{\nabla^2\tilde{Q}-\frac{1}{D}\frac{\partial\tilde{Q}}{\partial t}\right\}=0. \qquad (6.2)$$

In providing this new way to understand the quantum potential, Einstein's objection against both the completeness claim of the orthodox version and the Bohmian interpretation of quantum theory, using the example of a "particle in a box", was reiterated with a new perspective and finally resolved. This was done by proving that the corresponding quantum mechanical states exactly match classical analogues, i.e., said diffusion-wave fields. From these results, it is not hard to predict that the physics of diffusion waves may become a key to a deeper understanding of quantum mechanical nonlocality.

Alternative approaches to the orthodox view of quantum mechanics have accompanied the field since its very beginnings. Often they have not received much attention from a majority of physicists who were more interested in applications of the theory. As of today, however, this situation may change. It was shown here that the recently much elaborated fields of non-equilibrium thermodynamics and of diffusion-wave fields are essential for a causal understanding of the sub-quantum physics from which quantum theory is assumed to emerge. It is very encouraging to see that both fields mentioned are developing also, if not primarily, as new *experimental* disciplines. This makes it rather plausible that approaches like the one presented here may interest also a wider group of physicists. After all, there is now a very good prospect for fruitful mutual stimulations between the fields of non-equilibrium thermodynamics, diffusion waves, and quantum physics.